# A computed 95% confidence interval does cover the true value with probability 0.95 – epistemically interpreted

Dan Hedlin
Dept. of statistics
Stockholm University
106 91 Stockholm
Sweden

**Abstract**

Suppose the lifetime of a large sample of batteries in routine use is measured. A confidence interval is computed to $394 \pm 1.96 \cdot 4.6$ days. The standard interpretation is that if we repeatedly draw samples and compute confidence intervals, about 95% of the intervals will cover the unknown true lifetime. What can be said about the particular interval $394 \pm 1.96 \cdot 4.6$ has not been clear. We clarify this by using an epistemic interpretation of probability. The conclusion is that a realised (computed) confidence interval covers the parameter with the probability given by the confidence level is a valid statement, unless there are relevant and recognisable subsets of the sample.

Keywords and phrases: interpretation of confidence interval, epistemic probability, frequentist inference, recognisable subsets.

## 1. Introduction

Burnham, Lafta, Doocy and Roberts (2006) estimated the excess mortality after the invasion of the US-led coalition of Iraq in March 2003. Their estimate was 654,965 excess Iraqi deaths as a consequence of the war with the 95 percent confidence interval 392,979–942,636. The estimate was based on a probability sample with 1,849 responding households. The study was debated, sometimes animatedly. Marker (2008) reviews the methods of the study. He also provides some references to for example newspaper articles that scrutinised and criticised the study.

With this motivating example in mind, we focus on the interpretation of the confidence interval in a frequency context. The method of confidence intervals is usually viewed as a process that results in a relative frequency. We are now interested in the interpretation of the realised (computed) confidence interval. One common interpretation starts with the fact that the true excess mortality, denoted by $\theta$, and the end points of the interval 392,979, 942,636 are three numbers. As such, $\theta$ is either inside the interval 392,979–942,636, or it is not. Or in other words, the probability that $\theta$ is inside the interval is either 0 or 1. With this view on a confidence interval I cannot see any practical use of the concept of confidence intervals, and, as a consequence, I would find it hard to gauge the accuracy of the estimate 654,965. Another common and fundamentally different interpretation is that the interval does say something about the plausible error due to sampling. For example that the confidence interval provides a set of plausible values of $\theta$. But then the question is: what do we mean by 'plausible' here? Despite the lack of clarity the use of confidence intervals is ubiquitous. As an example of how confidence intervals may be reported in practice, see Sturgis et al. (2018).

After Jerzy Neyman's reading of his paper (Neyman, 1934) before the Royal Statistical Society, Arthur Bowley famously called the confidence interval a 'confidence trick' and went on to say: 'I think we are in the position of knowing that *either* an improbable event has occurred *or* the proportion in the population is within the limits. To balance these things we must make an estimate and form a judgment as to the likelihood of the proportion in the universe' (Bowley, 1934, p. 609).

Neyman focuses in his rejoinder (p. 623-625) first on Bowley's comment on the likelihood (chance) that the parameter is in the interval and notes that to say something about the probability of a parameter calls for Bayesian analysis. Then Neyman goes on to focus on Bowley's first comment, that all we know from an estimated confidence interval is either that something has occurred or that it has not occurred. Although Neyman acknowledges that Bowley is correct in saying that there is not much to say about the probability of a realised confidence interval, he states that his *method* (my italics) of confidence intervals should have a frequency interpretation.

Salsburg (2001, p. 122) believes that what is missing in the method of confidence intervals is the answer to this question: 'probability of what?' Salsburg captures the difficulty in one striking sentence: 'Bowley's problem with this procedure is one that has bedevilled the idea of confidence bounds since then'.

It is clear from Neyman's rejoinder that he views the method of confidence interval as a process by which a statement about the relative frequency is obtained. Neyman (1977, p. 118-119) gives two examples of statements that are results of incorrectly substituting realised values for the random variables in expression (1), which is given below: $P(1 \leq 5 \leq 3|\theta = 5) = 0.95$ and $P(1 \leq 2 \leq 3|\theta = 2) = 0.95$. Neyman (1977, p. 119) even spells out RELATIVE FREQUENCY in capital letters. Still, Salsburg's question is there: $P(1 \leq \theta \leq 3) = 0.95$, probability of what?

It is embarrassing that the concept of confidence interval, which is so important and so often used for the interpretation of statistical estimates, is still unclear 90 years after Neyman (1934) introduced both the concept of and the term confidence interval. I shall give a number of examples of various interpretations and vague statements. However, I believe the solution to the ambiguity and elusive clarity is quite simple: we should view a computed (realised) confidence interval in the light of an epistemic interpretation of probability while paying attention to the Fisherian concept of relevant and recognisable subsets.

Episteme is explained in the Oxford English Dictionary as 'scientific knowledge, a system of understanding; spec. (Foucault's term for) the body of ideas which shape the perception of knowledge in a particular period' with the ancient Greek origin meaning knowledge, understanding, skill and scientific knowledge. An anecdote illustrates the notion of epistemic probability. I decided to start a talk on philosophy of probability in a theatre-like hall by throwing a dice on stage. I took a good look at the outcome and asked the conferencier: 'what's the probability the outcome is 3?' One sixth he said (a tad anxiously). Then the next speaker, who the organisers had put in the first row right in front of the stage, said that to him the probability was one fourth. From his position he could see that the side of the dice had one single spot, and assuming that the dice was of standard construction, he could deduce that the far side must have six spots. So only four possible outcomes remained to him. Thus we had three probabilities, 1/6 (based on no knowledge), ¼ (based on partial knowledge) and 1 (because I knew it). Note that our probabilities do not have to be viewed as subjective probabilities. Everyone did a logical assessment based on facts (+ one assumption).

The idea to view confidence intervals in the light of epistemic probability is not new. Schweder (2018) suggests that. Pawitan, Lee and Lee (2023) investigate the issue if and under what conditions a nominal coverage probability applies to a realised confidence interval. They define the confidence level $\alpha$ to have the property 'epistemic confidence' if 'it is protected from the Dutch book' (see Vineberg, 2012, for the concept of Dutch book), assuming common interest and flow of information among people interested in the confidence interval at hand. What it basically means is that $\gamma = 1 - \alpha$ must follow the laws of probability and that there is an intersubjective agreement of the value of $\gamma$ (Gillies, 2000, p. 170-171). One condition is that there are no relevant subsets (we return to this in Sec. 6). The reasoning of Pawitan et al. (2023) is likelihood-based, in contrast to the current paper.

**2. Definition of a confidence interval**

We focus for simplicity on random sampling from finite populations and confidence sets that are in the form of a single interval. Let $\theta$ be a real-valued parameter of interest, for example the finite population average $\theta = \bar{y}_U = \sum_{k \in U} Y_k / N$, where $Y_1, Y_2, \ldots, Y_N$ are random variables and $U$ is the population with $N$ units. The population vector is $\mathbf{Y} = (Y_1, Y_2, \ldots, Y_N)'$. Let $\mathbf{Y}_s$ be the sample vector. A number of subsets of $U$ are possible samples through choice of sampling method. In the context here, it suffices to say that there are $B$ possible samples, $B$ being a finite number. Define a confidence interval as

$$P\big(L(\mathbf{Y}_s) \leq \theta \leq U(\mathbf{Y}_s)\big) = 1 - \alpha \tag{1}$$

where $L(\mathbf{Y}_s)$ and $U(\mathbf{Y}_s)$ are the random lower and upper confidence limits and $1 - \alpha$ is the confidence level. For simplicity we consider only $\alpha = 0.95$. We denote the confidence interval by $C(\mathbf{Y}_s) = \big(L(\mathbf{Y}_s), U(\mathbf{Y}_s)\big)$. A sample is collected, $\mathbf{y}_s$ is observed and $C(\mathbf{Y}_s)$ is estimated with $C(\mathbf{y}_s)$. We shall say that $C(\mathbf{y}_s)$ is the realised value of $C(\mathbf{Y}_s)$, and use the terms realised, estimated and computed interchangeably. If (1) is true, what is the probability in (2)? Is it meaningful to talk about such a probability? We assume that the sample is large; the focus here is on the interpretation of (2).

$$P\big(L(\mathbf{y}_s) \leq \theta \leq U(\mathbf{y}_s)\big) = 1 - \alpha \tag{2}$$

In model-based frequentist inference, the random sample is conditioned on and the limits in (1) are functions of the random variable $\mathbf{Y}_s$. In design-based survey sampling frequentist inference, moments are taken over the random sample $s$. A 95% confidence interval $CI(s)$ is an interval such that the probability that $CI(s)$ covers the unknown finite-population parameter $\theta$ is $P(CI(s) \ni \theta) = 0.95$. See Särndal, Swensson and Wretman (1992, p. 55). The sampling design makes $B$ samples possible to be realised. If the sample is large for all $B$ samples, approximately 95% of those would cover $\theta$, see Särndal et al. (1992) and references therein. My arguments do not depend on whether we consider model-based or design-based inference.

**3. Interpretations of confidence intervals in the literature**

A realised confidence interval, like the one in (2), is usually interpreted with its long-run relative frequency in mind, as Example 1 illustrates. The texts in Examples 1-2 and 4-11 are close to the source, although not necessarily direct quotations unless indicated. My interest in these example focuses on how authors approach (2).

**Example 1**
Thompson (1997, p. 144) conveys the idea behind classical sampling inference by way of an example. In an urn with $N$ balls there are $M$ white ones. A without-replacement simple random sample $s$ of size $n$ is going to be drawn. The random number of white balls in $s$ is $m_s$. The sample is drawn; it contains $m_s^0$ white balls. With $N = 100, n = 10, m_s^0 = 4$, Thompson estimates a 95% confidence interval for $M$ to [14,72] and goes on to say: 'Such intervals have the property that if the sampling procedure is repeated again and again, the long-run relative frequency of non-coverage will approximate [5%]'. (Continued in Example 9).

**Example 2**
A sample is drawn in order to estimate $\theta$, and a confidence is estimated. In their section 9.2.4, with the heading Bayesian intervals, Casella and Berger (2002) state that within classical statistics the parameter $\theta$ is fixed and the random confidence interval is also fixed after having been realised. So, as neither $\theta$ nor the realised interval moves, $\theta$ is either in the interval or it is outside. The probability that $\theta$ is inside is either 0 or 1.

Next example is taken from my own work as a statistician (modified and anonymised for simplicity and confidentiality).

**Example 3**
An entrepreneur in the health sector invoices a large hospital for a number of treatments of patients. The hospital takes a random sample of invoices and inspects the associated health records that the entrepreneur has written for each treatment. The hospital estimates that the entrepreneur has incorrectly invoiced $10,000 \pm 1,000$ money units where $(9,000, 11,000)$ is a 99% confidence interval.

They also compute a 95% confidence interval. The hospital sues the entrepreneur with the aim of salvaging at least the worth of the lower bound of the 99% interval. If the hospital had conducted a census, and all invoices had passed through the legal system, a total of $\theta$ money units would have been found to be incorrectly invoiced. However, following the interpretation of a confidence interval in Example 1 and in particular Example 2, we cannot say much about the probability that the interval (9,000, 11,000) contains $\theta$. So, is the confidence interval useless in this type of situation? And if it is useless here, in what situation would it be useful?

The next few examples seem to express that interpretation of a confidence interval is a matter of *confidence* or even hope.

**Example 4**
Cochran (1977, p. 12). If a probability sample of batteries in routine use in a large factory shows an average life of 394 days, with a standard error 4.6 days, the chances are 99 in 100 that the average life in the population of batteries lies between $394 \pm 2.58 \cdot 4.6$ days.

One may be forgiven if one suspects that Cochran dodges the term 'probability' in 'chances are'.

**Example 5**
Wooldridge (2013, p. 138-139). If random samples were obtained over and over again, with $L(\mathbf{Y}_s)$ and $U(\mathbf{Y}_s)$ computed each time, then the (unknown) population value $\theta$ would lie in the interval $C(\mathbf{Y}_s)$ for 95% of the samples. Unfortunately, for the single sample that we use to construct the confidence interval, we do not know whether $\theta$ is actually contained in the interval. We hope we have obtained a sample that is one of the 95% of all samples where the interval estimate contains $\theta$, but we have no guarantee.

**Example 6**
Agresti (2002, p. 411). The sample proportions of approval of a prime minister's performance are 0.59 and 0.55 in the first and second survey. Agresti suggests an estimator for the variance of dependent proportions. Using this, he computes the confidence interval for the difference in approval rating as (-0.06, -0.02) and formulates his conclusion in the following way: 'The approval rating appears to have dropped between 2 and 6 [percentage units]'.

However, in the next example, Foreman (1991) ventures to use the term probability.

**Example 7**
Foreman (1991, p. 42). Suppose the turnover of a business is estimated as ten million and that the standard error of that estimate is half a million. 'Then, on the supposition that the estimate is unbiased and that its sampling distribution approximates a normal distribution, it may be stated with 95% confidence (or probability $p = 0.95$) that the actual turnover… is within the interval' $10 \cdot 10^6 \pm 1.96 \cdot 0.5 \cdot 10^6$.

Cox and Hinkley (1974) use in the next example the word 'consistent' (in the colloquial meaning) but also seem to express some aversion in their use of the word 'legalistic'.

**Example 8**
Cox and Hinkley (1974, p. 209) describe the procedure to construct a confidence interval as something 'physical', which in principle could be checked by experiments. Nowadays we often make simulation studies where we draw a large number of samples and compute a coverage probability based on variance estimates. Such a simulation does provide a 'physical' interpretation of a confidence interval. Cox and Hinkley recommend that a system of confidence intervals is computed, in which the confidence level is varied, as opposed to selecting one confidence level in advance and compute only one interval. On p. 209 they say:

'In many ways a system of confidence limits calculated from data corresponds to treating the parameter as if, in the light of the data, it is a random variable $\Theta$ with cumulative distribution function implicitly defined by $P(\Theta \leq t^\alpha) = 1 - \alpha$', where $t^\alpha$ is the upper confidence limit in a one-sided interval.

'The distinction, emphasized in the standard treatments of confidence intervals, is that in the hypothetical repetitions that give physical meaning to $P(X \leq x)$, where $X$ is a random variable, we regard $x$ as a fixed arbitrary constant; more generally we consider the probability that $X$ falls in some *fixed* set. On the other hand, in [a computed confidence interval] it is $t^\alpha$ that varies in hypothetical repetitions, not $\theta$: hence it is not legitimate to treat $\theta$ as a random variable. Nevertheless, for many purposes the distinction appears to be a rather legalistic one'.

On page 214 Cox and Hinkley (1974) describe the procedure to construct a confidence interval from an acceptance region and write that the procedure 'suggests that we can regard confidence limits as specifying those parameter values consistent with the data at some level'. One reason for data not being consistent with the probability statement (1) is that (1) may not hold conditionally on ancillary statistics.

**Example 9** (continued from Example 1)
The realised confidence interval is 'a reasonable expression of inference about $M$. In intuitive terms, the computed 95% confidence interval for $M$ is compatible with the inference that, while we would guess that $M$ is in some interval around 40, we would be surprised to find it as low as 13 or as high as 73.' (Thompson, 1997, p. 144). However, as Thompson points out, an important difference between balls in an urn and, for example, an official statistics or a social survey is that the units in the latter are to a larger extent distinguishable. Individuals in a social survey may have labels (civic number, address) and auxiliary variables. The computed confidence interval may not be consistent with prior information or belief about the population (see examples in Section 6 below).

I interpret 'consistent' in Example 8 as 'given the data, there are no reasons not to believe that the computed confidence interval covers the parameter with a probability approximately equal to 0.95'. Example 9 provides one reason why we would not trust a realised confidence interval. We return to this in Section 6.

Like Cox and Hinkley in Example 8, Cramér (1956) seems to come close to saying that the probability by which a computed confidence interval covers $\theta$ is 0.95.

**Example 10**
Cramér (1956, p. 183). Suppose we conduct a series of experiments in which the parameters $\theta$, $\sigma$ and the number of units $n$ vary arbitrarily from experiment to experiment. For each experiment we compute a confidence interval and make the statement that $\theta$ is between the confidence limits. These limits will in general vary from experiment to experiment. The probability that our statement is correct is in each experiment 95%. If we consistently apply this rule, we would expect that 5% of our statements are false. The computed confidence interval gives us therefore a rule for estimation of $\theta$ that works with a constant failure risk of 5%.

Example 2 contrasts in an interesting way with the following example. Examples 2 and 11, it seems to me, capture the ambivalence in the interpretation of a confidence interval.

**Example 11**
Casella and Berger (2002, p. 421) : '...the confidence interval, the set in the parameter space with *plausible* values of $\mu$...' (my italics).

## 4. Theories of probability

Four broad theories (or interpretations) of probability are often mentioned (Gillies, 2000):
1. Logical. The degree of rational belief. The anecdote with the dice in Sec. 1 is one example.
2. Subjective, the degree of belief of a particular individual, and intersubjective, a consensus in a group of people
3. Frequency
4. Propensity, which is wide and multifaceted class of theories. Popper (1959) advocated one version.

Theories 1 and 2 are often called epistemic or subjective probabilities. The latter term is unfortunate as the same term refers both to the group and to one member of the group. Theories 3 and 4 are referred to as objective, scientific or aleatory (Gillies, 2000, p. 20). The division between epistemic and objective probabilities makes many see probability as being Janus-faced.

Daston (1988, p. 197) gives this reason as to why no distinction was made between epistemic and objective probability in the past: 'Experience generated belief and probability by the repeated correlation of sensations which the mind reproduced in associations of ideas. The more constant and frequent the observed correlation, the stronger the mental association, which in turn intensified probability and belief. Hence, the objective probabilities of experience and the subjective probabilities of belief were, in a well-ordered mind, mirror images of one another.'

One way to understand subjective probability is to regard it as a degree of belief. It is 'a causal property … which we can express vaguely as the extent of which we are prepared to act on it' (Frank Ramsay, cited in Gillies, 2000, p. 54). Suppose you are tempted to bet on a game between teams A and B. If you bet $u_2$ money units on 'A win', and A come out victorious, then you obtain $u_1$ money units plus your own bet. A fair bet is characterised by a zero expected profit, that is, $u_1 p - u_2(1-p) = 0$ where $p$ is the probability for team A win, implying that the fair bet is characterised by $p = u_2 / (u_1 + u_2)$ and hence that the fair (statistical) odds is $p/(1-p) = u_2 / u_1$. Imagine that three odds, $odds_1, odds_2$ and $odds_3$, are offered to person D, who does not know the real value of $p$. Suppose $odds_1$ is higher than the fair one, that is, $p_1/(1-p_1) > p/(1-p)$, where $p_1$ is the probability that corresponds to $odds_1$. If person D bets on $odds_1$, he effectively believes that A win with probability $p_1$ and he will lose money in the long run. As discussed by Daston (1988), person D will with experience learn that he plays too hard (ignoring that there are psychological gambling issues). Suppose now that $odds_2$ and $odds_3$ are $p_2/(1-p_2) < p_3/(1-p_3) < p/(1-p)$. If he bets on $odds_2$ but not on $odds_3$, he is too risk averse. The rational choice is to play on an odds which is as close as possible to the fair one, but not higher. This in turn implies that D is prepared to act on a subjected probability up to $p$, but not further. See also Gillies (2000, p. 55) for a trick of how to tease the subjective probability out of a person who is inclined to 'cheat' to make an expected profit. The ratio $p = u_2 / (u_1 + u_2)$ is called 'betting quotient'.

As an argument why considering betting to be relevant to other decisions, Frank Ramsay writes: 'Whenever we go to the station we are betting that a train will really run, and if we had not sufficient degree of belief in this we should decline the bet and stay at home.' (Cited in Gillies, 2000, p. 54).

Also, think of evidence *e* as observations or some assumed state of affairs (the trains are usually on time, and there is this morning no reason to believe otherwise), and hypothesis *h* as a statement about the unknown state of affairs (the train will run). If a person's subjective probability of *h* given *e* is high, then that would give that person good reasons for expecting the unknown facts described by *h* (Carnarp, 1950, p. 164). Looking back to Example 4, Cochran (1977) seems to come close to a subjective probability.

Not everybody accepts that a *thrown* dice has a probability of a certain outcome; if it is a 3, then it is what it is, whether we have observed the outcome or not. However, the Royal Society (2020, p. 12) disagree: 'It is a common misapprehension that probabilities can only be used for future events with some randomness. While it is true that an event has either happened or not, many statisticians will feel that it is reasonable to assign probabilities to our personal uncertainty about unknown facts.' The document of the Royal Society addresses the use of probability in legal contexts. What is the probability that a fragment of glass, which was found on a suspect's clothing, came from a particular broken window? If you do not accept that a thrown dice has a probability for a certain outcome (which has already happened), then you cannot reasonably accept that there is a probability that the fragment came from a specific window, because then you would rather think 'either it did, or it did not'.

Fisher (1958, p. 386-387) clearly had an epistemic interpretation in mind when he wrote: 'Probability is… a well specified state of logical uncertainty…. Let us consider any uncertain event. A child is going to be born…[and being interested in the sex of the child] we are in a state of uncertainty'. A parent finds out that in 51 per cent of births, the baby is a boy. Fisher goes on to say: 'he [the dad]…informs us that the probability of a boy is 51 per cent, having made reference to this measurable reference set as the basis of his statement.'

## 5. The main argument and three counterarguments

Consider two experiments. In the first one you have 100 balls, 95 of which are blue and 5 are red, and one ball is going to be selected at random. That is, $P(blue) = 0.95$. In the second experiment the balls are behind a screen and one ball has already been selected at random. Most statisticians (and indeed most non-statisticians) would now say the probability that the colour of the ball is blue is 0.95. Now suppose the set of balls are $B$ confidence intervals and 95% of them have the property $E$, where $E$ is the event that the confidence interval covers $\theta$. One confidence interval has been selected randomly. There are two, or possibly three, reasons why one would not accept that a confidence interval which has already 'been selected' (realised) has $E$ with probability 0.95. The counterarguments are:
1. There is no such a thing as epistemic probabilities
2. We might gain some knowledge by inspecting the realised sample that would make us distrust the confidence interval
3. We should not, or cannot, mix subjective probabilities with frequencies

I have said enough about counterargument 1. Next I shall discuss counterargument 3, and in Section 6 discuss the interesting counterargument 2.

A subjective probability may serve as an estimate of a relative frequency. A rigorous argument is given by Carnarp (1950, §41), which is now summarised briefly, although recast to the confidence interval problem. Denote the possible confidence intervals by $C^{(i)}(\mathbf{y}_s)$, $i = 1,2,...,B$, and let $E_i$ be the event that $C^{(i)}(\mathbf{y}_s)$ covers $\theta$. Suppose all $B$ confidence intervals have been realised. Let $r$ be the proportion of the $B$ confidence intervals which cover $\theta$. Person A, with no knowledge of the property $E_i$ of any of the realised confidence intervals, believes that a fair odds for each $E_i$ is $u_2 / u_1$, so her subjective probability is the betting quotient $q = u_2/(u_1 + u_2)$. If A places a bet on each one of the $B$ confidence intervals, her expected gain is $ru_1 - (1-r)u_2 = r(u_1 + u_2) - (u_1 + u_2)q = (u_1 + u_2)(r - q)$. Thus, her betting quotient need be $q = r$ for it to be fair (for the expected gain to be zero). In other words, her best subjective probability is also her estimate of the relative frequency $r$ of the $C^{(i)}(\mathbf{y}_s)$ that cover $\theta$. Hence, a tenable subjective probability is also an estimate of the relative frequency. This argument seems to be what Daston (1988), cited above, discusses.

## 6. Relevant and recognisable subsets

Now we turn to counterargument 2. A statistician may, and often does, identify relevant and recognisable subsets (a term borrowed from R.A. Fisher) in the sample. Three examples follow.

The proportions of men and women in the sample may be very different from known population proportions. If gender is associated with the study variable then the sample mean of a simple random sample is in the presence of sample imbalance conditionally biased. In the Burnham et al. (2006) study there were more violent deaths among men than women, so if the sample had contained an excess of women, the estimate would have been biased low (unless successfully adjusted for with estimation, e.g. poststratification).

Brewer (2002, p. 271) tells the story when he as a young statistician discovered that the two primary sampling units in one stratum both were railway junctions and therefore were atypical in terms of their workforce composition. Brewer's supervisor told him to discard one of them and select another one randomly. That is, the supervisor asked Brewer to tamper with a realised probability sample. But, although the young Brewer had misgivings, the experienced Brewer believed that 'it was better to have a reasonably well-balanced sample that a pure probability sample that was obviously badly balanced' (p. 271).

When inspecting a computed confidence interval you might as an experience practitioner see that it is very wide, and identify the reason: one large value of the study variable.

Fisher (1959, p. 23) requires two properties: 'No subset can be *recognized* having a different probability. Such subsets must always exist; it is required that no one of them shall be recognizable. This is a postulate of ignorance'. Women form the subset in the first example. In the second example the subset is all non-railway junctions, which was conspicuously empty. In the third example the large value constitutes the subset.

The three examples illustrate reasons why there may be a gap between our prior knowledge and the computed confidence interval. In all three examples we cannot have full confidence in the confidence interval. Really, we should be reluctant to publish such confidence intervals (unless, again, we have been able adjust for the issues, with estimation).

To define 'recognisable and relevant subsets' we first introduce a vector of auxiliary variables $\mathbf{x}_k$ for unit $k$, which could be for example sex and age. Suppose $P(L(Y_s) \leq \theta \leq U(Y_s)) = 1 - \alpha$ where $Y_s$ is a stochastic vector of sample values, but $P(L(Y_s) \leq \theta \leq U(Y_s)|R(\mathbf{y}_s, f(\mathbf{x}_s), f(\mathbf{x}_U))) \neq 1 - \alpha$, where $R(\cdot,\cdot,\cdot)$ is a statistic, which may depend on sample data $\mathbf{y}_s$, $\mathbf{x}_s$ auxiliary variables in the sample and $\mathbf{x}_U$, auxiliary variables in the population (or stratum or cluster depending on application). The subscript $U$ indicates that $\mathbf{x}_U$ contains other information about the population than what could be deduced from the sample. The statistic depends on auxiliary variables through a function $f(\cdot)$. If there is such a statistic $R(\cdot,\cdot,\cdot)$, we say that it induces 'relevant and recognisable subsets', similarly to Buehler (1959) and Pawitan et al. (2023). In the Burnham et al's (2006) study (in our thought-experiment with an identified excess of women) and in Brewer's (2002) survey, $x_k$ is a binary that takes value 1 if unit $k$ is a woman or a non-railway junction, $f(\mathbf{x}_s)$ is the proportion of women or non-railway junctions in the sample, respectively. In the study of excess mortality in Iraq, $f(\mathbf{x}_U)$ is the proportion of women in the country. In Brewer's survey $f(\mathbf{x}_U)$ is the known proportion of railway junctions in the stratum, and the two induced subsets are the whole sample and the empty set. In neither case $f(\mathbf{x}_U)$ needed to be known with high accuracy for one to realise that the sample was imbalanced. In the third example, where the sample contained a large value that made the confidence interval very wide, the relation between $f(\mathbf{x}_U)$ and $f(\mathbf{x}_s)$ shows that the sample is imbalanced. Even if there was in this third example no auxiliary variable, the statistician may still see that the sample is imbalanced by making an experienced judgement of what the population mean $\bar{\mathbf{y}}_U$ should be compared to the observed $\bar{\mathbf{y}}_s$.

Pawitan et al. (2023) proves that, under some conditions, there cannot be any relevant subsets. Here it is interesting to note Fisher's (1959) statement: "Such subsets must always exist". The reason why Pawitan et al. (2023) are able to prove their theorem is that they work with likelihoods, for example, that a sample is drawn from a gamma distribution. Fisher on the other hand is thinking of a real-life population. Suppose you have interviewed people in Iraq. In that case there may be unobserved variables that, if known, would make you realise that the sample is imbalanced. Recall Thompson's remark in Example 9 above that there is a difference between official statistics or a social survey and 'balls in an urn'.

Pawitan et al. go on to give conditions for $1 - \alpha$ being epistemic.

## 7. Prosecutor's fallacy in relation to this discussion

Suppose we are about to construct a 95% confidence interval from the null hypothesis $H_0: \theta = \theta_0$. The acceptance region $A(\theta_0)$ is given by sample data $\mathbf{y}_s$ that lead to a computed estimate $\hat{\theta}(\mathbf{y}_s)$ for which

$$\theta_0 - 1.96\sqrt{V(\hat{\theta}(\mathbf{y}_s))} \leq \hat{\theta}(\mathbf{y}_s) \leq \theta_0 + 1.96\sqrt{V(\hat{\theta}(\mathbf{y}_s))} \tag{3}$$

We may invert (3) into a confidence interval in the usual way to

$$\hat{\theta}(\mathbf{y}_s) - 1.96\sqrt{V(\hat{\theta}(\mathbf{y}_s))} \leq \theta \leq \hat{\theta}(\mathbf{y}_s) + 1.96\sqrt{V(\hat{\theta}(\mathbf{y}_s))}. \tag{4}$$

Since (3) defines an acceptance region $A(\theta_0)$, $P(\hat{\theta}(\mathbf{Y}_s) \in A(\theta_0) | \theta = \theta_0) = 0.95$. However, if we now claim that $P(\theta = \theta_0 | \hat{\theta}(\mathbf{Y}_s) \in A(\theta_0)) = 0.95$, then 'prosecutor's fallacy' may come up in the mind of a statistician (e.g. Taylor, 2018).

What is the difference between prosecutor's fallacy and the probability referred to in Example 7? They are probabilities of different events. In the fallacy it is the illegitimate interpretation of the parameter $\theta$ being a random number. In Example 7 and other examples above it is the probability (epistemically interpreted) that we have obtained sample $i$ and confidence interval $C^{(i)}(\mathbf{y}_s)$. This may be the answer to Salsburg's (2001, p. 122) question 'probability of what?'.

## 8. Discussion and conclusion

There are disparate interpretations of confidence interval in the literature, as shown in Examples 1-11. These examples suggest that statisticians have an awkward relationship to the interpretation of confidence intervals.

There are situations where the computed confidence interval is not convincing. Many of these are well known to practitioners. In situations such as those given in Sec. 6, the computed confidence interval is not consistent with our prior knowledge or belief. Quite simply: there is something wrong with the confidence interval.

Unless there are relevant and recognisable subsets (as defined in Sec. 6), it is not incorrect to say that computed 95% confidence interval covers the parameter with probability 0.95, that is, (2) is valid. This epistemic interpretation of that probability has the additional advantage that it defines the term 'confidence'.

One final thought about the 90 years of bewilderment since Neyman (1934): it might be a widely held view in the statistical community that a probability should *either* be seen as subjective or viewed in a frequency context that has created a barrier that made us fail to see the fairly simple solution.